# The Biological Origin of Linguistic Diversity

*Andrea Baronchelli[1], Nick Chater[2], Romualdo Pastor-Satorras[3],*
and *Morten H. Christiansen[4,5] ***

[1] Laboratory for the Modeling of Biological and Socio-technical Systems, Northeastern University, Boston MA 02115 USA

[2] Behavioural Science Group, Warwick Business School, University of Warwick, Coventry, CV4 7AL, UK

[3] Departament de Física i Enginyeria Nuclear, Universitat Politècnica de Catalunya, Campus Nord B4, E-08034 Barcelona, Spain

[4] Department of Psychology, Cornell University, Uris Hall, Ithaca, NY 14853, USA

[5] Santa Fe Institute, 1399 Hyde Park Road, Santa Fe, NM 87501, USA

[*] Email: christiansen@cornell.edu

**Abstract**

In contrast with animal communication systems, diversity is characteristic of almost every aspect of human language. Languages variously employ tones, clicks, or manual signs to signal differences in meaning; some languages lack the noun-verb distinction (e.g., Straits Salish), whereas others have a proliferation of fine-grained syntactic categories (e.g., Tzeltal); and some languages do without morphology (e.g., Mandarin), while others pack a whole sentence into a single word (e.g., Cayuga). A challenge for evolutionary biology is to reconcile the diversity of languages with the high degree of biological uniformity of their speakers. Here, we model processes of language change and geographical dispersion, and find a general consistent pressure for flexible learning, irrespective of the language being spoken. This pressure arises because flexible learners can best cope with observed high rates of linguistic change associated with divergent cultural evolution following human migration. Thus, rather than genetic adaptations for specific aspects of language, such as recursion, the coevolution of genes and fast-changing linguistic structure has produced a biological basis fine-tuned to linguistic diversity. Only biological adaptations for flexible learning combined with cultural evolution can explain how each child has the potential to learn any human language.

# Introduction

Natural communication systems differ widely across species in both complexity and form, ranging from the quorum-sensing chemical signals of bacteria [1], to the colour displays of cuttlefish [2], the waggle dance of honeybees [3], and the alarm calls of vervet monkeys [4]. Crucially, though, within a given species, biology severely restricts variability in the core components of the communicative system [5], even in those with geographical dialects (e.g., in oscine songbirds [6]). In contrast, the estimated 6-8,000 human languages exhibit remarkable variation across all fundamental building blocks from phonology and morphology to syntax and semantics [7]. This diversity makes human language unique among animal communications systems. Yet the biological basis for language, like other animal communication systems, appears largely uniform across the species [8]: children appear equally able to learn any of the world's languages, given appropriate linguistic experience. For example, aboriginal people in Australia diverged genetically from the ancestors of modern European populations at least 40,000 years ago [9], but readily learn English. This poses a challenge for evolutionary biology: How can the *diversity of human language* be reconciled with its presumed *uniform biological basis*?

Linguistic diversity and the biological basis of language have traditionally been treated separately, with the nature and origin of the latter being the focus of much debate. One influential proposal argues in favour of a special-purpose biological language system by analogy to the visual system [10, 11, 12, 13]. Just as vision is crucial in navigating the physical environment, language is fundamental to navigating our social environment. Other scientists have proposed that language instead relies on domain-general neural mechanisms evolved for other purposes [14, 15, 16]. Just as reading relies on neural mechanisms that predate the emergence of writing [17], so perhaps language has evolved to rely on pre-existing brain systems. However, there is more agreement about linguistic diversity, which is typically attributed to divergent cultural evolution following human migration [9]. As small groups of hunter-gatherers dispersed geographically, first within and later beyond Africa [18], their languages also diverged [19].

Here, we present a theoretical model of the relationship between linguistic diversity and the biological basis for language. Importantly, the model assigns an important role to linguistic change, which has been extraordinarily rapid during historical times; e.g., the entire Indo-European language group diverged from a common source in less than 10,000 years [20]. Through numerical simulations we determine the circumstances under which the diversity of human language can be reconciled with a largely uniform biological basis enabling each child to learn any language. First, we explore the consequences of an initially stable population splitting into two geographically separate groups. Second, we look at the possibility that such groups may not be separated, but continue to interact to varying degrees. Third, we consider the possibility that linguistic principles are not entirely unconstrained, but are partly determined by pre-existing genetic biases. Fourth, we investigate the possibility of a linguistic "snowball effect," whereby linguistic change was originally slow—allowing for the evolution of a genetically specified protolanguage—but gradually increased across generations. In each of these cases, we find that the evolution of a genetic predisposition to accommodate rapid cultural evolution of linguistic structure is key to reconciling the diversity of human language with a largely uniform biological basis for learning language.

## Methods

**The Model**

A population of $N$ agents speaks a language consisting of $L$ principles, $P_1,.. P_L$. Each individual is endowed with a set of $L$ genes, $G_1,.. G_L$ each one coding for the ability to learn the corresponding principle. A linguistic principle is a binary variable that can assume one of two values: $+L$ or $-L$. Each gene has three alleles, $+G, -G$ and $?G$: the first two encode a bias towards learning the $+L$ and $-L$ principle, respectively, and the third is neutral. Learning occurs through a trial and error procedure. The allele at a given locus determines the learning bias towards the corresponding linguistic principle. If locus $i$ is occupied by allele $+G$, the individual guesses that the linguistic principle $P_i$ is $+L$ with a probability $p>0.5$ and that it is $-L$ with probability *(1-p)*. The expected number of trials to guess the right principle is therefore *1/p* if the allele favours that principle and *1/(1-p)* if it favours the opposite one. The

"ideal" genome for learning of the target language consists of alleles favouring the principles of that language. The closer a genome approaches this ideal, the faster learning occurs—with no learning required in the ideal case—thus implementing a genetic endowment specific to language [21, 22, 23, 24, 25]. Neutral alleles, by contrast, allow for maximal flexibility in learning, not tied to specific linguistic principles.

Following previous work suggesting that rapid learning language contributes to individual reproductive success [26], we define the fitness of an individual to be inversely proportional to the total time $T$ spent by that individual to learn the language. Specifically, $T = \sum_{i=1}^{L} t_i$, where $t_i$ is the number of trials the individual requires to guess principle $i$. At each generation, a fraction $f$ of the population, corresponding to the $fN$ individuals with the highest fitness, is allowed to reproduce. Pairs of agents are then randomly chosen and produce a single offspring by sexual recombination: for each locus of the "child", one of the two parents is randomly chosen and the allele for the corresponding locus is copied. With probability $m$, moreover, each allele can undergo random mutation.

The language also changes across generations, with each principle subject to mutation with a probability $l$. This random change of language can be viewed as a possible consequence of cultural pressures that may, for example, drive languages of separate groups apart, so that the languages can function as a hard-to-imitate marker of group identity [27].

Typical values of the parameters are *N=100, L=20, p=0.95, m=0.01* and *f=0.5* (see [28] for discussion of the robustness of the model against changes in these parameter setting).

**Population Splitting**

After a certain number of generations (typically *500* or *1000* in our simulations and generally after the onset of a steady state), the population is split in two new subpopulations of size *N'*. These subpopulations inherit all the parameters set at the beginning for the prior population, as well as its language, but then evolve independently. Throughout, we set *N'=N*, to rule out possible effects of population size (hence, strictly speaking, the population is cloned).

**Divergence Measures**

When a population reaches a steady state, it is split into two "geographically separated" subpopulations that evolve independently. We measure the linguistic as well as genetic divergence between these two populations and determine which initial conditions yield realistic scenarios concerning language origins. Given populations *A* and *B*, their *linguistic* divergence $D_L(A, B)$ is computed as the normalized Hamming distance between the two languages; i.e., $D_L(A, B) = H(A, B)/L$, where $H(A, B)$ simply counts the number of corresponding principles which are set on different values. Formally, $D_L(A, B)$ evolves as a function of the number of generations *t* as (see the Appendix for the derivation of Eq. 1):

$$D_L(A,B) = \frac{1 - e^{-4lt}}{2} \tag{1}$$

Similarly, *genetic* divergence $D_G(A, B)$ quantifies the degree to which alleles are shared across two populations $A$ and $B$, averaged over $L$ genes. In general, we consider that two populations are similar if they share a large fraction of their genetic material. To deal with the fact that alleles have three variants, we need a simple generalization of Hamming distance to measure similarity between "genomes." For each locus $i$, we determine the frequency $n_x$ of each allele, where $x = ?G, +G$ and $-G$, in both populations $A$ and $B$. The overlap, or "similarity", on the allele $x$ is then given by the *minimum* of the two, $\min[n_x(A), n_x(B)]$. The total similarity $s_i$ at locus $i$ reads

$$s_i(A,B) = \min[n^i_{+G}(A), n^i_{+G}(B)] + \min[n^i_{-G}(A), n^i_{-G}(B)] + \min[n^i_{?G}(A), n^i_{?G}(B)].$$

Hence, $s_i = 0$ if the two populations are completely misaligned, say because in one of them all the individuals have the $?G$ allele while in the other all individuals have the $+G$ variant, and $s_i = 1$ if they are identical. The normalized similarity between the two populations is therefore $S(A,B) = \frac{1}{L} \sum_{i=1}^{L} s_i(A,B)$. The genetic divergence is then defined simply as the complementary $D_G(A,B) = 1 - S(A,B)$.

# Results

**Population Divergence**

We first consider the evolution of genes and language in a single population that splits into two separate subpopulations. Because our simulations incorporate both biological adaptation of learners as well as cultural evolution of languages, this allows us to test whether a special-purpose language system could have co-evolved with language itself [21, 22, 23, 24, 25].

Figure 1*a* shows that, if the rate of language change *l* is small, genomes adapt to the language change in each population. Thus the genes of the two populations drift apart, yielding very different biological bases for language with strong genetic biases (i.e., almost no neutral genes). Figure 1*b* illustrates that by contrast if *l* is large, neutral genes are favoured in both populations. This is because the language is a fast-moving target, and committing to a biased allele to capture the current language will become counterproductive, when the language changes. So, while languages diverge, the genes in the different populations remain stable, primarily consisting of neutral genes. The insert in figure 1*a* shows the interplay between the rates of genetic mutation and linguistic change. Below a critical value of *l*, genes adapt to linguistic change (the fraction of neutral alleles is negligible); otherwise, language-specific adaptation does not occur (neutral alleles predominate).

Our results exhibit two patterns. If language changes rapidly it becomes a moving target, and neutral genes are favoured in both populations. Conversely, if language changes slowly, two isolated subpopulations that originally spoke the same language will diverge linguistically and subsequently biologically through genetic assimilation to the diverging languages. Only the first pattern captures the observed combination of linguistic diversity and a largely uniform biological basis for language, arguing against the emergence of a special-purpose language system.

**Interaction between Populations**

Might a less complete population splitting yield different results? Hunter-gatherers typically have local contact, especially by marriage, so that people frequently need to learn the languages of more than one group [29]. Could the exposure to a more complex, multi-lingual environment lead to the evolution of a special-purpose language system? We investigate these questions relating to interactions between populations in a second set of simulations.

After the population splitting, as above, contact between the two subpopulations is modelled by letting an individual's fitness be determined by the ability not just to learn the language of their own group, but also the other group's language. Specifically, each individual has a probability $C$ of having to learn the language of the other population. The case $C=0$ corresponds therefore to the usual setting of completely isolated groups, as before; $C=0.5$ describes two populations whose individuals are randomly exposed to one of two independent languages. Although each agent only has to learn a single language, our simulation corresponds functionally to a situation in which an individual must to have the appropriate genetic basis for learning both languages.

Figure 2 shows the impact of a multi-lingual environment on genetic divergence. We only consider slow linguistic change because, as we have seen, at large $l$ neutral genes predominate and no special-purpose language system can evolve. The results indicate that small values of $C$ do not alter the picture observed for complete isolation; and where $C$ increases, neutral $?G$ alleles predominate for both groups: again, no genetic assimilation to specific aspects of language occurs.

**Divergent Gene-Language Coevolution**

The current model misses a crucial constraint, by assuming that language change is random. But language might be partially shaped by the genes of its speakers. Could such reciprocal influence of genes on language be crucial to explaining how a special-purpose language system might coevolve with language? In a third set of simulations, we introduce a parameter $g$ that implements a genetic pressure on language change at each generation. Thus, at each generation, with probability $g$ the linguistic principle at locus $i$ is deterministically set to be maximally learnable by the population, i.e., to mirror the most frequent non-neutral genetic allele in the corresponding location. Otherwise, with probability $1 - g$, the linguistic principle under consideration mutates, as before, with probability $l$ or remains unchanged with probability $1 − l$. Similar to the previous simulations, the mother population splits after a certain number of generations and the two new populations evolve independently.

Figure 3*ab* illustrates that with small *l*, low *g* yields a scenario in which genes and languages remain constant across generations, even after population splitting. This stasis is not compatible with observed linguistic diversity. When *l* is large, as in figure 3*cd*, and genetic influence is low, neutral alleles predominate and populations remain genetically similar, as before. As *g* increases, genetic influence reduces language change; language becomes a stable target for genetic assimilation. Consequently, the biased +*G* and –*G* alleles dominate, but genes diverge between subpopulations. For larger values of *g*, the influence of genes on language eliminates linguistic (and subsequent genetic) change. None of these regimes produce the combination of linguistic diversity and genetic uniformity observed across the world today. Rather, this pattern only emerges for low *g* and high *l*, yielding a predominance of neutral alleles inconsistent with the idea of a special-purpose language system.

**An Early Protolanguage?**

So far, we have shown that a uniform special-purpose language system could not have coevolved with fast cultural evolution of language, even if linguistic change is driven by genetic pressures. But perhaps early language change was slow. After all, the archaeological record indicates very slow cultural innovation in, for example, tool use, until 40,000-50,000 years ago [30]. Perhaps a genetically-based special-purpose language system coevolved with an initially slowly-changing language—a 'protolanguage' [31—and these genes were conserved through later periods of increased linguistic change? We therefore simulated the effects of initially slow, but accelerating, language change across generations.

In the final set of simulations, the linguistic mutation rate *l* was not held constant, but increased linearly with generations. More precisely, at the beginning of the simulation we set *l* = 0. Then, the value of *l* is increased at each generation by a value of *δl* = *0.1/M*, where *M* is the total number of generations, so that at the end of the simulation *l* = *0.1*. As usual, after *M/2* generations the population splits and two new subpopulations keep evolving independently. In the cases presented here, *M* = *2000*.

Figure 4 shows that in a single population, it is adaptive to genetically align with a stable linguistic environment. But as the speed of linguistic change increases, the number of neutral alleles increases. This continues after the population splits: languages diverge and the genes of both subpopulations are predominantly neutral—undoing the initial genetic adaptation to the initial language. The results suggest that even if a uniform special-purpose language system could adapt to a putatively fixed protolanguage, it would be eliminated in favour of general learning strategies, as languages later became more labile. This argues against a "Prometheus" scenario [32] in which a single mutation (or very few) gave rise to the language faculty in an early human ancestor, whose descendants then dispersed across the globe. Our results further imply that current languages are unlikely to carry within them significant "linguistic fossils" [33] of a purported initial protolanguage.

**Discussion**

Our results indicate that humans have evolved a biological adaptation specifically for keeping up with the cultural evolution of language, instead of a special-purpose linguistic system analogous to the visual system. While vision has developed over hundreds of millions of years across many species, language has arisen only in hominids, over hundreds of thousands of years [9]. Importantly, whereas the visual world is relatively invariant across time and space, each language user must deal with an ever-changing linguistic environment, created by other language-users. Although there is evidence of gene-culture coevolution (e.g., lactose tolerance appears to have coevolved with dairying [34]), a special-purpose language system would have had to coevolve with a constantly changing linguistic environment. Yet, the geographical spread of human populations creates linguistically isolated populations, with gradually diverging languages, and hence diverging selectional pressures. Thus, just as Darwin's finches adapted to their local island environments, coevolution with language would lead to geographically separated human populations each with a distinct special-purpose language systems coevolved with its local linguistic environment. Thus, genetic populations should be adapted to their *own* language families; but this is not empirically observed. Thus, our results suggest, instead, that humans have evolved a flexible learning system to follow rapid linguistic change. This evolutionary outcome is robust even when separate populations continue to intermix, when language change is partly determined by genetic factors, and when initially slow rates of linguistic change are assumed.

To reconcile linguistic diversity with a largely uniform biological basis for language, our results point to an evolved genetic predisposition to accommodate to the continual cultural evolution of language. Only then can we explain the observed pattern of (i) great variety across the world's languages; (ii) that genetic origins have little or no impact on ease with which people learn a given language. We speculate that the cultural evolution of language may further have recruited pre-existing brain systems to facilitate its use [14, 16], just as reading and writing appear to rely on prior neural substrates [17]. Constraints on these 'recycled' neural systems may accordingly have shaped the cultural evolution of language without promoting additional language-specific genetic changes [16, 28, 35, 36]. Thus, linguistic diversity arises from an evolved genetic adaptation for cultural linguistic evolution, additionally shaped by non-linguistic constraints deriving from a largely uniform biological basis of general perceptual, cognitive, and pragmatic abilities that predate the emergence language.

More generally, our findings complement recent results from the application of computational phylogenetic methods to large databases of typological language information (see [37] for a review), indicating that patterns of word order correlations across language families are best understood in terms of lineage specific histories of cultural evolution rather than as the reflection of a special-purpose language system [38]. Such phylogenetic analyses, however, do not provide direct insight into the circumstances under which gene-language coevolutionary processes may give rise to the diversity of human language as investigated in our simulations. Thus, we advocate a two-pronged methodological approach to language evolution that combines insights from phylogenetic methods into historical processes relating to the diversification of languages with those from computational and mathematical modelling of the coevolutionary interplay between genes and language.

# Appendix

**Derivation of Equation 1**

Consider two languages, *A* and *B,* and a given linguistic principle. Given that the principle is a binary variable, either the two languages share the same value of the principle (i.e., they both have +L or -L) or they are misaligned (one of them has +L, and the other one –L). In the first case, we say they are in *s* state, and in the latter that they are in *d* state. The probability of a switch from the *s* state to the *d* state is twice the language mutation rate *l*, since it is produced every time that one of the two languages mutates. Calling *P(x)* the probability of having the two languages in state x, and considering that *P(d)=1-P(s)*, we have:

$$\frac{\partial P(d)}{\partial t} = 2l[P(s) - P(d)] = 2l[1 - 2P(d)].$$

The Hamming distance between the two languages and on the given principle is *H(A,B)=1* if the languages are in the *d* state and *H(A,B)=0* when they are in the *s* state. Therefore, the average values is exactly H(A,B)=P(d). Considering that right after a population split the two languages are in the *s* state, the above equation straightforwardly yields the solution:

$$P(d) = H(A,B) = \frac{1 - e^{-4lt}}{2},$$

which also describes the time evolution of $D_L(A, B)$ because linguistic principles evolve independently. It is worth noting that this argument relies on the assumed independence of the different linguistic principles. An alternate scenario in which some principles are correlated with other ones is out of the scope of the present paper, but likely worth future investigations.

# References

1. Miller MB, Bassler BL (2001) Quorum sensing in bacteria. Annu Rev Microbiol 55: 165-

2. Hanlon RT, Messenger JB (1988) Adaptive coloration in young cuttlefish (*Sepia officinalis L.*): the morphology and development of body patterns and their relation to behaviour. Phil Trans R Soc Lond B 320: 437-487.

3. Dyer FC (2002) The biology of the dance language. Annu Rev Entomol 47: 917-949.

4. Seyfarth RM, Cheney DL (1984) The natural vocalizations of non-human primates. Trends Neurosci 7: 66-73.

5. Hauser MD (1996) The evolution of communication. Cambridge, MA: MIT Press.

6. Gil D, Gahr M (2002) The honesty of bird song: multiple constraints for multiple traits. Trends Ecol Evol 17: 133-141.

7. Evans N, Levinson S (2009) The myth of language universals: language diversity and its importance for cognitive science. Behav Brain Sci 32: 429-492.

8. Pinker S (1994) The language instinct. New York: William Morrow.

9. Cavalli-Sforza LL, Feldman MW (2003) The application of molecular genetic approaches to the study of human evolution. Nat Genet 33: 266-275.

10. Chomsky N (1980) Rules and representations. Oxford: Blackwell.

11. Fodor J (1983) The modularity of mind. Cambridge. MA: MIT Press.

12. Maynard-Smith J, Szathmáry E (1997) Major transitions in evolution. New York: Oxford University Press.

13. Pinker S (1997) How the mind works. New York: WW Norton.

14. Deacon TW (1997) The symbolic species: the coevolution of language and the brain. New York: WW Norton.

15. Tomasello M (2008) Origins of human communication. Cambridge, MA: MIT Press.

16. Christiansen MH, Chater N (2008) Language as shaped by the brain. Behav Brain Sci 31: 489-558.

17. Dehaene S, Cohen L (2007) Cultural recycling of cortical maps. Neuron 56: 384-398.

18. Tishkoff SA, et al. (2009) The genetic structure and history of Africans and African Americans. Science 324:1035-1044.

19. Richerson PJ, Boyd R (2010) Why possibly language evolved. Biolinguistics 4: 289-306.

20. Gray RD, Atkinson QD (2003) Language-tree divergence times support the Anatolian theory of Indo-European origin. Nature 426: 435-439.

21. Dunbar RIM (2003) The origin and subsequent evolution of language. In: Christiansen MH, Kirby S, editors. Language evolution. New York: Oxford University Press. pp. 219-234.

22. Nowak MA, Komarova NL, Niyogi P (2001) Evolution of universal grammar. Science 291: 114-118.

23. Pinker S, Bloom P (1990) Natural selection and natural language. Behav Brain Sci 13: 707-784.

24. Számadó S, Szathmáry E (2006) Selective scenarios for the emergence of natural language. Trends Ecol Evol 21: 555-561.

25. Tooby J, Cosmides L (2005) Conceptual foundations of evolutionary psychology. In: Buss DM, editor. The handbook of evolutionary psychology. Hoboken, NJ: Wiley. pp 5-67.

26 Komarova NL, Nowak MA (2001) Natural selection of the critical period for language acquisition. Proc R Soc B 268: 1189-1196.

27 Boyd R, Richerson PJ (1987) The evolution of ethnic markers. Cultural Anthropology 2: 65–79.

28 Chater N, Reali F, Christiansen MH (2009) Restrictions on biological adaptation in language evolution. Proc Natl Acad Sci USA 106: 1015-1020.

29. Lee RB, Daly RH (1999) The Cambridge encyclopedia of hunters and gatherers. Cambridge: Cambridge University Press.

30. Mithen SJ (1996) The prehistory of the mind: a search for the origins of art, science and religion. London: Thames & Hudson.

31 Bickerton D (1995) Language and human behavior. London: UCL Press.

32 Chomsky N (2010) Some simple evo devo theses: How true might they be for language? In: Larson RK, Déprez V, Yamakido H, editors. The evolution of human language. Cambridge: Cambridge University Press. pp. 45-62.

33 Jackendoff R (1999) Possible stages in the evolution of the language capacity. Trends Cogn Sci 3: 272-279.

34. Beja-Pereira A, *et al*. (2003) Gene-culture coevolution between cattle milk protein genes and human lactase genes. Nat Gen 35: 311-313.

35. Smith K, Kirby S. (2008) Cultural evolution: implications for understanding the human language faculty and its evolution. Phil Trans R Soc B 363: 3591-3603.

36. Griffiths TL, Kalish ML, Lewandowsky S (2008) Theoretical and experimental evidence for the impact of inductive biases on cultural evolution. Phil Trans R Soc B 363: 3503-3514.

37 Levinson SC, Gray RD (2012) Tools from evolutionary biology shed new light on the diversification of languages. Trends Cogn Sci 16: 167-173.

38 Dunn M, Greenhill SJ, Levinson SC, Gray RD (2011) Evolved structure of language shows lineage-specific trends in word-order universals. Nature 473 79-82.

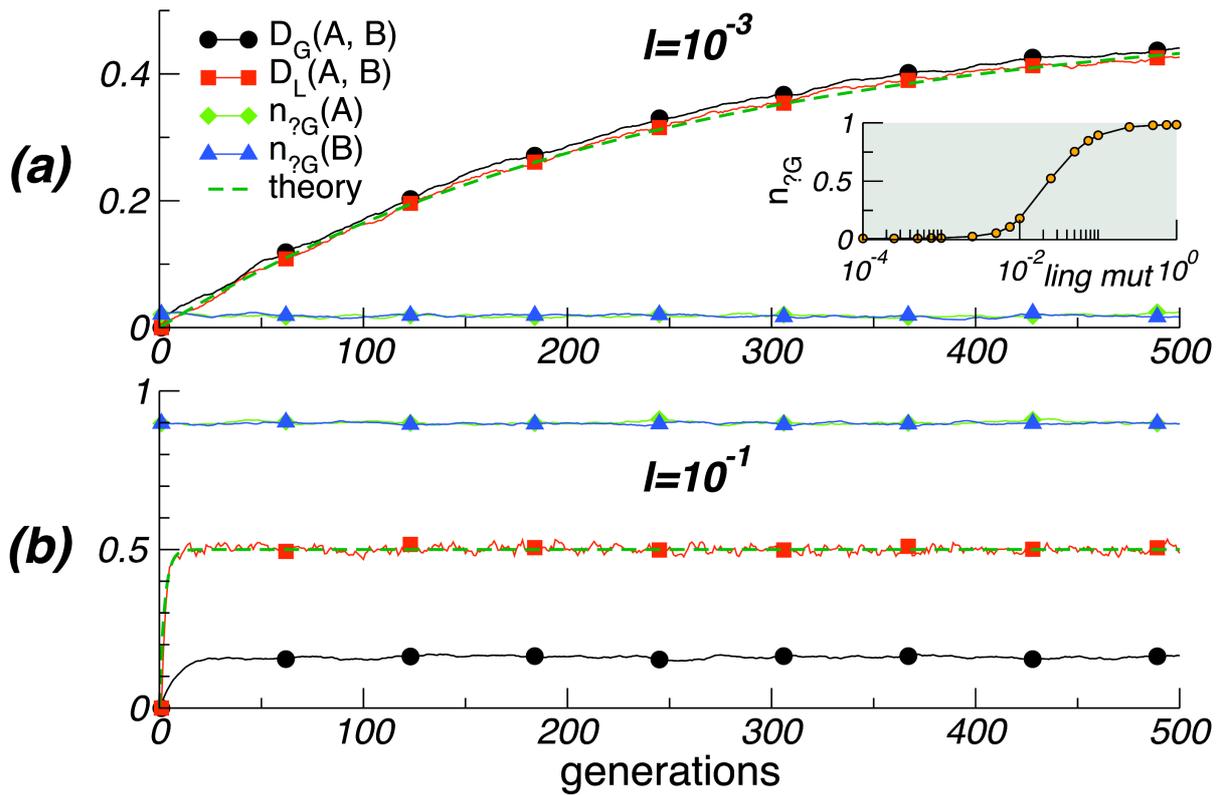

**Figure 1. Population divergence**. $D_G(A, B)$ and $D_L(A, B)$ measure the genetic and linguistic divergence of the two subpopulations (created at generation $0$), while $n_x$ measures the frequency of allele $x$. *(a)* When language change is slow, language provides a fixed target for the genes. As the two languages diverge, the genes for each subpopulation follow its 'local' language, and thus also diverge. Neutral alleles are eliminated except for occasional re-emergence through mutation. The insert shows the stationary fraction of neutral alleles as a function of the rate of linguistic change. *(b)* With faster linguistic change, languages diverge immediately but the populations remain genetically similar, dominated by neutral alleles. The rate of linguistic change is derived through eq. 1. Parameters are: $N = 100$, $m = 0.01$, $L = 20$, $f = 0.5$; the initial fractions of alleles in the original population are $n_{?g}(t=0)=0.5$, $n_{+G}(t=0)=n_{-G}(t=0)=0.25$. Curves are averaged over 100 runs.

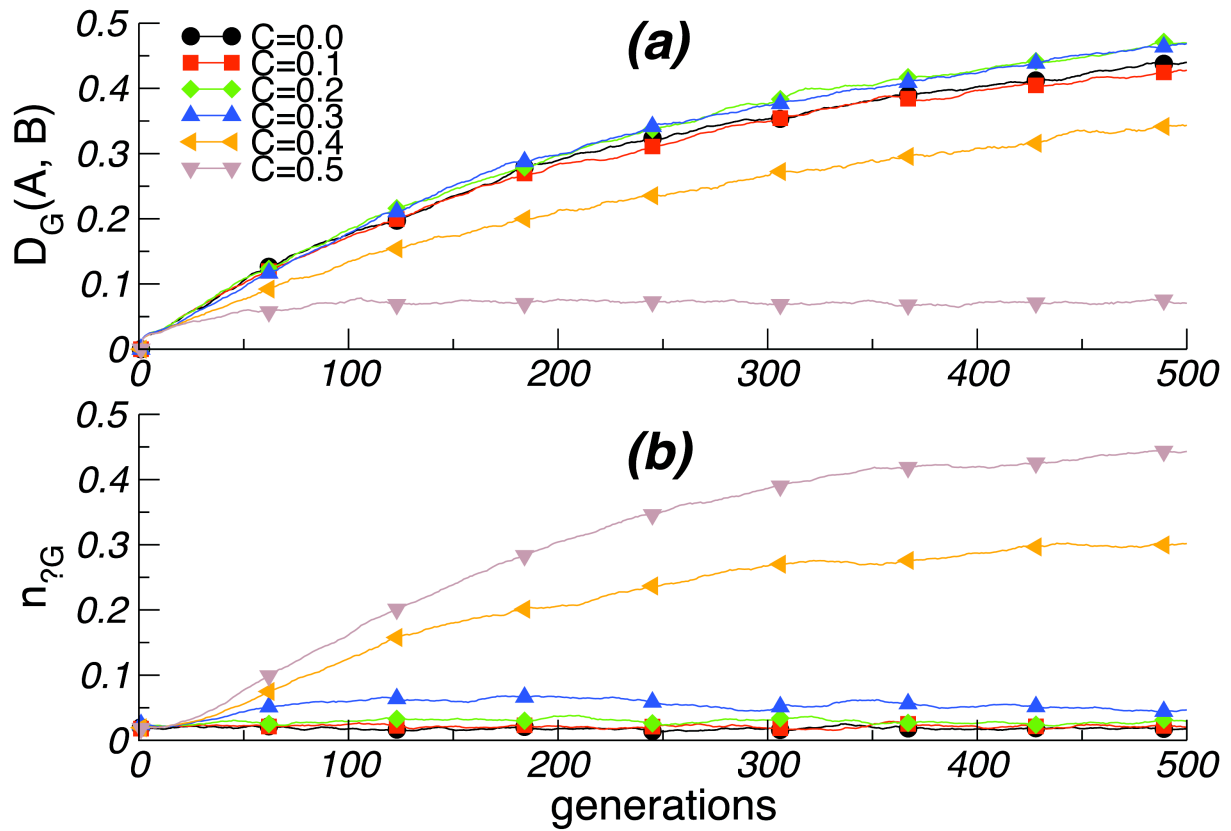

**Figure 2. Interaction between subpopulations.** *(a)* $C$ controls the probability that the fitness is determined by an individual's ability to learn the language of the other population. We assume slow language change ($l=10^{-3}$; other parameters as in figure 1). When $C=0$, language is a stable target for the genes and the two subpopulations diverge genetically, with few neutral alleles. *(b)* As $C$ increases, neutral genes predominate, and the subpopulations are genetically similar. The panel shows a single subpopulation.

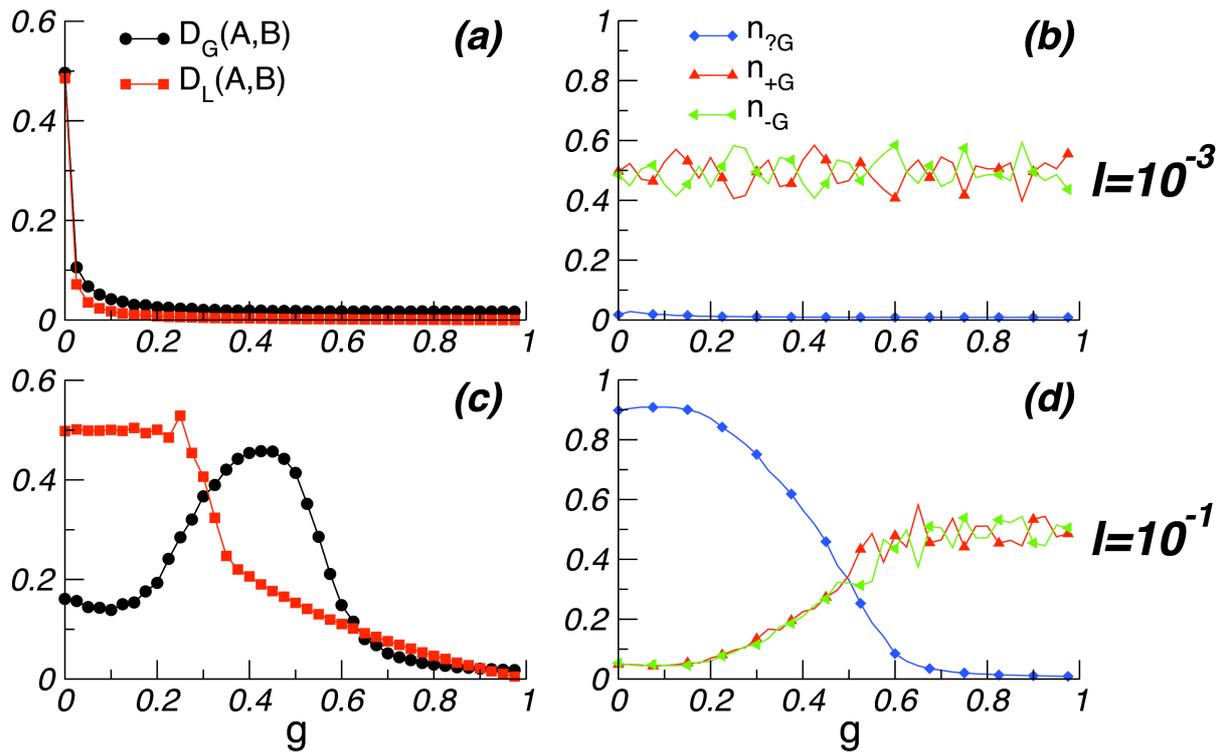

**Figure 3. Influence of genes on languages.** *(a)-(b)* For slow language change, small *g* triggers stasis: neither languages nor genes evolve. *(c)-(d)* When language changes rapidly, only small values of *g* are compatible with both rapidly diverging languages and small genetic divergence between the populations. Other parameters as in figure 1.

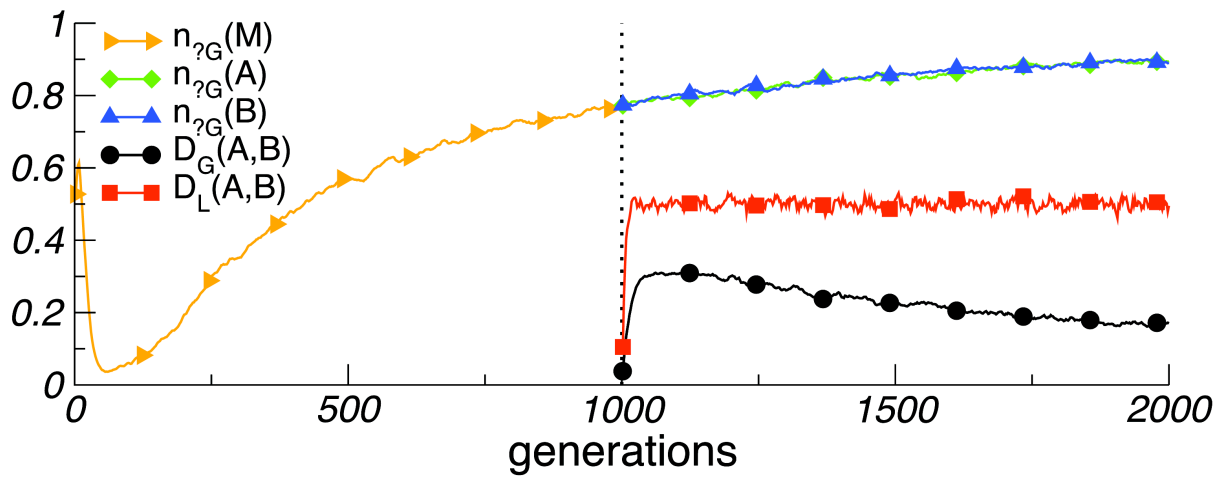

**Figure 4. Early protolanguage.** Here the probability $l$ that a linguistic principle changes increases gradually across generations by a small amount $\delta l$. At generation *1000*, the population, $M$, splits. The number of neutral alleles continues to grow in the two subpopulations, $A$ and $B$, which therefore become increasingly genetically similar (their genetic divergence decreases). Yet their languages diverge rapidly. Parameters: $\delta l = 0.1/W$, with $W=2000$, $l(t=0)=0$, $l(t=2000)=0.1$; others as in figure 1.